\documentclass[10pt,letterpaper]{article}
\usepackage{opex3}
\usepackage{fixltx2e}
\usepackage{epstopdf}

\begin{document}


\title{Waveguide-Coupled Photonic Crystal Cavity for Quantum Dot Spin Readout}
\author{R.J. Coles$^{1*}$, N. Prtljaga$^{1}$, B. Royall$^{1}$, I.J. Luxmoore$^{2}$, A.M. Fox$^{1}$ and M.S. Skolnick$^{1}$}

\address{$^{1}$Department of Physics and Astronomy, University of Sheffield, S3 7RH, United Kingdom \\
$^{2}$College of Engineering, Mathematics and Physical Sciences, University of Exeter, EX4 4QF, United Kingdom}

\email{$^{*}$rjcoles1@sheffield.ac.uk}


\begin{abstract}
We present a waveguide-coupled photonic crystal H1 cavity structure in which the orthogonal dipole modes couple to spatially separated photonic crystal waveguides. Coupling of each cavity mode to its respective waveguide with equal efficiency is achieved by adjusting the position and orientation of the waveguides. The behavior of the optimized device is experimentally verified for where the cavity mode splitting is larger and smaller than the cavity mode linewidth. In both cases, coupled Q-factors up to 1600 and contrast ratios up to 10 are achieved. This design may allow for spin state readout of a self-assembled quantum dot positioned at the cavity center or function as an ultra-fast optical switch operating at the single photon level.
\end{abstract}




\section{Introduction}

Scalable all-optical quantum information processing (QIP) has been shown to be possible using only single-photon sources, linear optical elements and single-photon detectors \cite{O'Brien2009,Nielsen2004,Knill2001}. The two-level spin system of a self-assembled quantum dot single-photon source is one of the leading candidates for a static qubit implementation, \cite{Divincenzo2000} with long dephasing times \cite{Borri2001} and possibility of optical coherent control \cite{Bonadeo1998}. On chip integration using this solid state implementation requires a static to flying qubit (spin-photon) interface to exchange quantum information between different static nodes \cite{Divincenzo2000}. Recent demonstrations of the entanglement between a QD and single photon \cite{Gao2012,DeGreve2012} and the mapping of QD spin states to path-encoded photons \cite{Luxmoore2013} are important milestones in the development of QD-based solid-state QIP.

For many purposes, path encoding with indistinguishable photons is desired \cite{Imamog1999} which could be achieved by inclusion of a unpolarized optical cavity \cite{Thijssen2012a}. The two TE dipole modes form a Poincar\'{e}-like sphere with states $\alpha\left|X\right\rangle\pm\beta\left|Y\right\rangle$ which have a one-to-one correspondence to the in-plane QD spin states $\alpha\left|x\right\rangle\pm\beta\left|y\right\rangle$, where $\alpha$ and $\beta$ are complex \cite{Thijssen2012a}. In addition, due to the low mode volume (V) of these dipole modes, the H1 cavity possesses one of the highest Q/V ratios of any PhC cavity \cite{Takagi2012}: the resulting high degree of spontaneous emission enhancement has been shown to provide indistinguishable single photon emission \cite{Laurent2005}, strong coupling \cite{Ota2009} and entangled photon pairs \cite{Larque2009}. On-chip coupling of the hexapole \cite{Kim2004} and quadrupole \cite{Yu2012} modes of the H1 cavity to waveguides has been investigated previously, but a demonstration of the selective coupling of the dipole modes to separate waveguides remains to be demonstrated as required for in-plane transmission of spin \cite{Luxmoore2013}. We propose a scheme whereby the Poincar\'{e}-like states of the H1 cavity are mapped into two separate propagating photon channels, allowing information encoding the QD spin state to be transferred to the waveguides via the cavity.

In this paper we use FDTD simualations to design a waveguide-coupled H1 device which exhibits selective coupling of each of the two dipole modes to its respective PhC waveguide, and then demonstrate its operation using ensemble QD photoluminescence (PL) measurements for near-degenerate and non-degenerate cavities. We use the QD ensemble as an internal light source to characterize the cavity modes.

\section{Device Design \& Optimisation}
\subsection{Design Principles}

\begin{figure}[h]
\centering
\includegraphics[width=8.8cm]{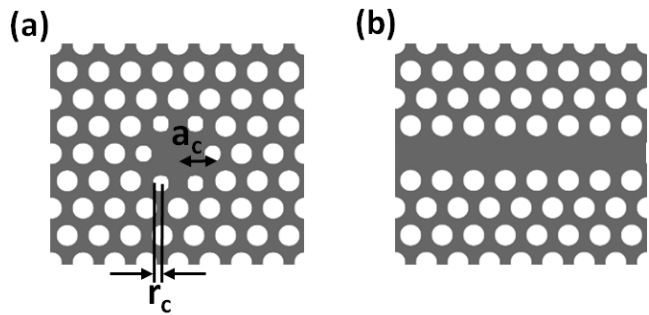}
\caption{(a) Optimized H1 cavity structure with $r_{c}=0.91r$ and $a_{c}=1.09a$. (b) Line-defect (W1) photonic crystal waveguide.}
\label{bare_cavity}
\end{figure}

\begin{figure}[h]
\centering
\includegraphics[width=12.2cm]{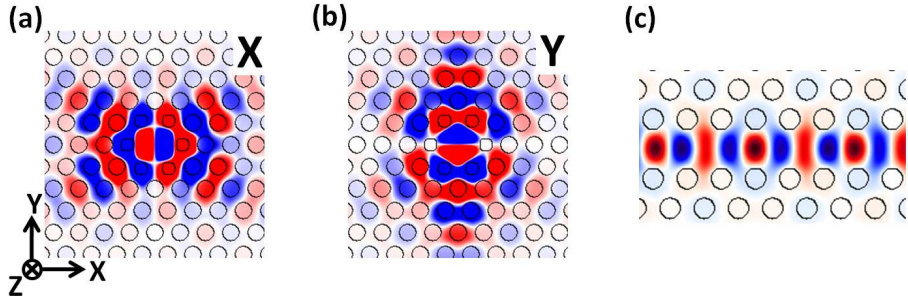}
\caption{(a) \& (b) Normalized H\textsubscript{z} near-field amplitudes of the (a) X \& (b) Y-dipole modes. The modes are labeled according to the orientation of the H\textsubscript{z} dipole. A linear red-white-blue color scale is applied to represent fields up to 50\% of the maximum value. Values above this have a saturated red or blue color. (c) Normalized H\textsubscript{z} field amplitude of the odd parity guided mode of the W1 waveguide, using a full range linear red-white-blue color scale.}
\label{cav_modes}
\end{figure}

The optimization of the device was performed by using the finite-difference-time-domain (FDTD) computational method via the freely available software package MEEP \cite{meep}. The H1 cavity consists of an omitted air cylinder from a triangular-lattice photonic crystal (PhC) slab with lattice constant $a$, cylinder radius $r=0.31a$, slab thickness $h=0.71a$ and refractive index $n=3.4$. To maximize the cavity Q-factors, the six nearest-neighbor cylinders have reduced radii $r_{c}=0.91r$ and increased displacement $a_{c}=1.09a$, as shown in Fig. \ref{bare_cavity}(a), producing calculated Q factors for the dipole modes of $\sim30,000$ with a mode volume $V=0.39(\lambda/n)^{3}$ \cite{Shirane2007}. The near-field profiles of the dipole modes are shown in Fig. \ref{cav_modes}(a) \& (b) where the modes are labeled according to the orientation of the  H\textsubscript{z} dipole at the cavity center. Herein referred to as the X and Y-dipole modes, the X-dipole has an H\textsubscript{z} dipole along the x-axis and the Y-dipole along the y-axis. With careful geometric arrangement of two photonic crystal waveguides, we show that it is possible to selectively couple these cavity modes to the guided modes of two separate waveguides.

In the spectral region of the cavity modes, the linear defect (W1) waveguide shown in Fig. \ref{bare_cavity}(b) sustains a single, propagating TE mode \cite{Notomi2001}, the H\textsubscript{z} field profile of which is shown in Fig. \ref{cav_modes} (c). The cavity modes will couple to the waveguide provided that there is good spatial overlap and the cavity mode field symmetry matches that of the waveguide mode \cite{Kim2004,Faraon2007a,Alija2006,Martinez2005,Schwagmann2012}. 

The H\textsubscript{z} fields of the cavity modes in Fig.\ref{cav_modes}(a) \& (b) decay evanescently into the surrounding photonic crystal, exhibiting significant penetration into the PhC along the dipole axis and vanishingly small fields orthogonal to it. For the X-dipole mode the field symmetries match those of the waveguide mode in Fig.\ref{cav_modes}(c), both possessing odd parity in the y=0 plane and even parity in the x=0 plane, whilst the cavity mode field symmetries are opposite for the Y-dipole mode. Therefore, a W1 waveguide brought into close proximity to the cavity along the x-axis will couple well to the X-dipole but poorly to the Y-dipole. The same selection principle holds for coupling along the Y-axis, except the PhC lattice symmetry forbids a W1 waveguide along this axis. In this case, a waveguide at 30$^{\circ}$ to the vertical can be employed however, such that the waveguide terminates on a line along the y-axis that passes through the cavity center as shown in Fig.\ref{coup_eff_calc}(a). Along this line, the waveguide field overlap remains high for the Y-dipole and low for the X-dipole mode.

It should be noted that although the cavity modes also exhibit significant penetration depths at 45$^{\circ}$ to the mode axis, these are common to both modes and are therefore unsuitable for selective coupling.

\subsection{Optimisation Procedure}
\begin{figure}[h]
\centering
\includegraphics[width=8.8cm]{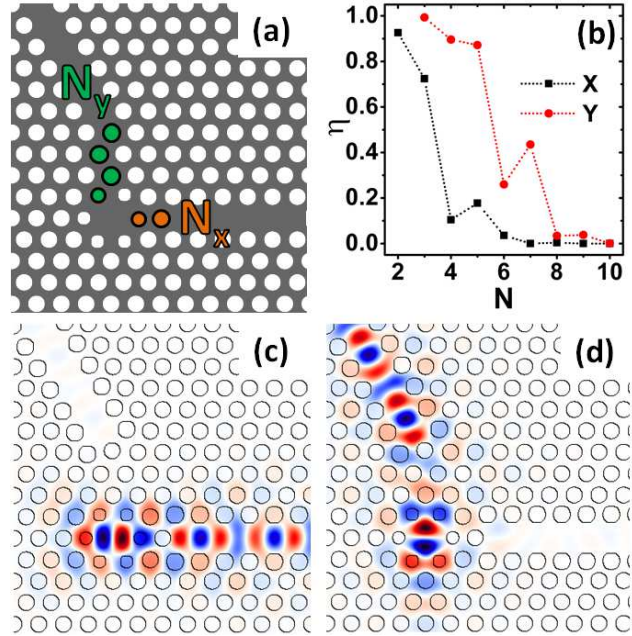}
\caption{(a) Coupled cavity-waveguide structure (b) Coupling efficiency between the cavity modes and waveguides as a function of the number of holes separating them. (c) \& (d) Normalized Hz field amplitudes of the coupled systems using $N_{x}=2, N_{y}=4$ for (c) X \& (d) Y, respectively. The color scale is the same as used in Fig. \ref{cav_modes}(c).}
\label{coup_eff_calc}
\end{figure}

For optimal device operation, the modes must couple to the waveguides with equally high efficiencies, whilst maintaining the highest Q-factor possible to provide maximum spontaneous emission enhancement to the QD.  The coupling efficiency was calculated for each mode as a function of the number of holes separating the cavity and waveguide (N\textsubscript{x} \& N\textsubscript{y} for the x \& y-waveguide respectively, as illustrated in Fig. \ref{coup_eff_calc}(a)). We define the x(y)-waveguide as that which principally couples to the X(Y)-dipole. The coupling efficiency is defined as \cite{Faraon2007a}

\begin{equation}\label{coup_eff_eq}
\eta(N_{x,y})\equiv\frac{Q_{wg}(N_{x,y})^{-1}}{Q_{c}(N_{x,y})^{-1}}=1-\frac{Q_{c}(N_{x,y})}{Q_{u}}
\end{equation}

\noindent where $Q_c(N_{x,y})$ and $Q_{u}$ are the Q-factors of the coupled and uncoupled cavity respectively and $Q_{wg}^{-1}$ is the loss rate into the waveguide, given by

\begin{equation}
Q_{wg}(N_{x,y})^{-1}=Q_{c}(N_{x,y})^{-1}-Q_{u}^{-1}
\end{equation}

The coupling efficiencies to each waveguide, calculated separately, are shown in Fig. \ref{coup_eff_calc} (b). As expected, the coupling efficiency decreases with an increase of the cavity-waveguide separation, due to a reduction in the evanescent tunneling. The exceptions of N\textsubscript{y}=5 and 7 for the Y-dipole are due to the path taken when adjusting the cavity-waveguide separation, resulting in the waveguide moving in and out of the evanescent tail. The mechanism is different for the X-dipole at N\textsubscript{x}=5, since the cavity and waveguide modes share the same axis: this phenomenon is attributed to fluctuations in the overlap integral of the two modes \cite{Faraon2007a,Chalcraft2011}.

When both waveguides were introduced, the coupling efficiencies for X(Y)-waveguide separations for 2(4) hole separation were found to be comparable at 89(93)\%. The H\textsubscript{z} field profiles of the coupled modes are shown in Fig. \ref{coup_eff_calc}(c) \& (d). H\textsubscript{z} fields were chosen so that both waveguide modes could be observed simultaneously and any cross talk would be evident by visual inspection. The cross talk coupling was calculated to be \textless 10\% for the separation values given above, highlighting that each waveguide has little perturbation on the other.

\begin{figure}[h]
\centering
\includegraphics[width=13.2cm]{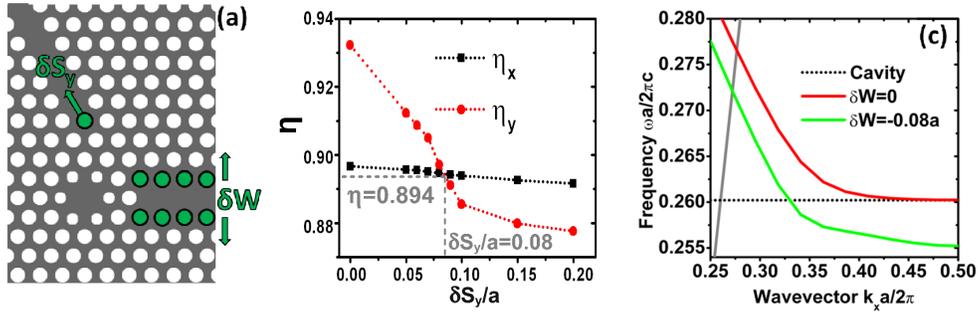}
\caption{(a) Schematic defining $\delta$Sy and $\delta$W. $\delta$W was applied to both waveguides, but is only illustrated on the X-waveguide for clarity. (b) Coupling efficiency of the cavity modes to their respective waveguides as the first hole in the Y waveguide is shifted. (c) Dispersion of waveguides as inner hole rows are displaced outward.  The dispersion curves were calculated using the frequency domain iterative eigensolver, MIT Photonic Bands (MPB) \cite{mpb}.}
\label{fine_tuning}
\end{figure}

Whilst discrete adjustment of the number of holes between cavity and waveguide produces comparable coupling efficiencies, a continuous adjustment is needed to equalize these values. This is achieved by displacing the first hole in the Y-waveguide along the waveguide by $\delta S_{y}$, as defined in Fig. \ref{fine_tuning}(a). This reduction in coupling efficiency for the Y-dipole equalizes both efficiencies at 89\% for $\delta S_{y}=0.08a$ as shown in Fig. \ref{fine_tuning}(b).

The W1 waveguide is known to produce slow-light phenomena near to the band edge \cite{Krauss2007}, leading to increased scattering losses \cite{Wasley2012} and spectral cut-off of the cavity modes \cite{Waks2005}. The resonant frequency of the cavity modes is coincident with the band edge of the waveguide dispersion as shown in Fig. \ref{fine_tuning}(c). To avoid this the waveguide mode is red-shifted by displacing the first row of holes perpendicular to the waveguide by $\delta W=0.08a$ (Fig. \ref{fine_tuning}(a)), ensuring the cavity mode is coincident with a higher group velocity region of the waveguide dispersion \cite{Yu2012,Kim2004} as shown in Fig. \ref{fine_tuning}(c). $\delta W$ was applied to both waveguides, but is only illustrated for on the X-waveguide in Fig. \ref{fine_tuning}(a) for clarity.

\section{Experimental Results}
\subsection{Experimental Arrangement}
\begin{figure}[h!]
\centering
\includegraphics[width=9.9cm]{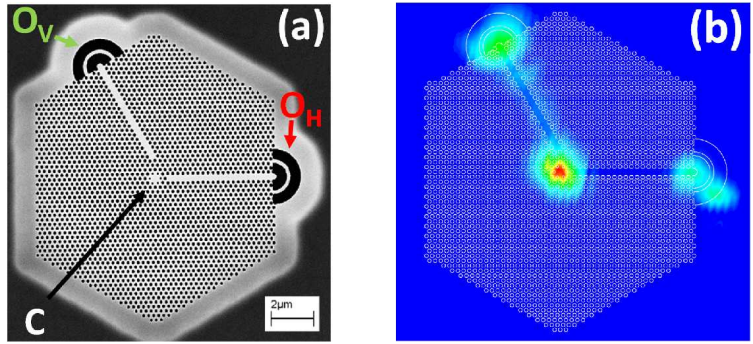}
\caption{(a) SEM image of fabricated device. O$_H$ \& O$_V$ denote the vertical and horizontal outcouplers respectively, C is the cavity. (b) PL map obtained from a raster scan of the excitation spot whilst keeping collection fixed at the cavity. The spectrometer was used to filter PL from the center wavelength of the cavity modes. A contour of the device structure is overlaid.}
\label{sem_pl_image}
\end{figure}

The samples used in this study were grown by molecular beam epitaxy (MBE) on undoped GaAs (100) wafers. The wafer consisted of a 140nm GaAs layer containing a layer of self-assembled InAs QDs at its center, above a 1$\mu$m thick sacrificial Al$_{0.6}$Ga$_{0.4}$As layer on an undoped GaAs substrate. The photonic crystal was patterned by electron beam lithography, followed by an inductively-coupled plasma etch to define the pattern into the GaAs membrane. The  Al$_{0.6}$Ga$_{0.4}$As layer was removed by an isotropic hydrofluoric acid etch to leave a free-standing air-clad GaAs slab. A scanning electron microscope image of the fabricated device is shown in Fig. \ref{sem_pl_image}(a). Semicircular $\lambda/2n$ air/GaAs grating outcouplers were added to the end of the waveguides to scatter light out of the device plane into the detection apparatus \cite{Faraon2008}.

Photoluminescence measurements were performed with a confocal scanning microscopy setup with the sample mounted in a liquid helium bath-cryostat at ~4.2K \cite{Grazioso2010}. The sample was excited with an 850nm CW Ti:Sapphire laser focussed to a spot of ~1$\mu$m by a 0.62NA objective lens. The QD PL was collected using the same objective before being filtered by a 900nm long pass filter and dispersed by a 0.55m single spectrometer onto a liquid nitrogen cooled charge-coupled device (CCD) camera or passed through additional slits and incident upon a fast avalanche photodiode (APD). A motorized scanning mirror was employed in both the excitation and detection paths which allows for spatially selective excitation and detection from the sample \cite{Wasley2012}. Fig. \ref{sem_pl_image}(b) shows a typical PL map obtained when the excitation spot is rastered across the device whilst the APD collects spectrally filtered PL at the cavity peak from a fixed collection spot over the cavity position: the cavity modes are efficiently coupled and transmitted by the waveguides. The different positions used for excitation and collection are defined in Fig. \ref{sem_pl_image}(a): O\textsubscript{H} and O\textsubscript{V} are the horizontal and vertical outcouplers, respectively and C is the cavity. We define a notation to identify these positions of excitation and collection spots as excitation/collection. For example, excitation of the cavity and collection from the horizontal outcoupler is denoted C/O\textsubscript{H}.

\subsection{Non-Degenerate Cavity}

The H1 cavity is highly sensitive to the symmetry of the surrounding photonic crystal. In  fabrication, random disorder commonly leads to reduction of the cavity symmetry and lifting of the degeneracy of the dipole modes. Schemes have been demonstrated to remedy this by adjusting the ellipticity of the holes in the photonic crystal \cite{Luxmoore2011} and applying uniaxial strain to the wafer \cite{Luxmoore2012}. We did not apply these techniques to our devices, however, as an average spectral splitting of $\sim$1.5nm facilitates mode identification, allowing spectral measurements to reveal the coupling behavior.

\begin{figure}[h]
\centering
\includegraphics[width=12.2cm]{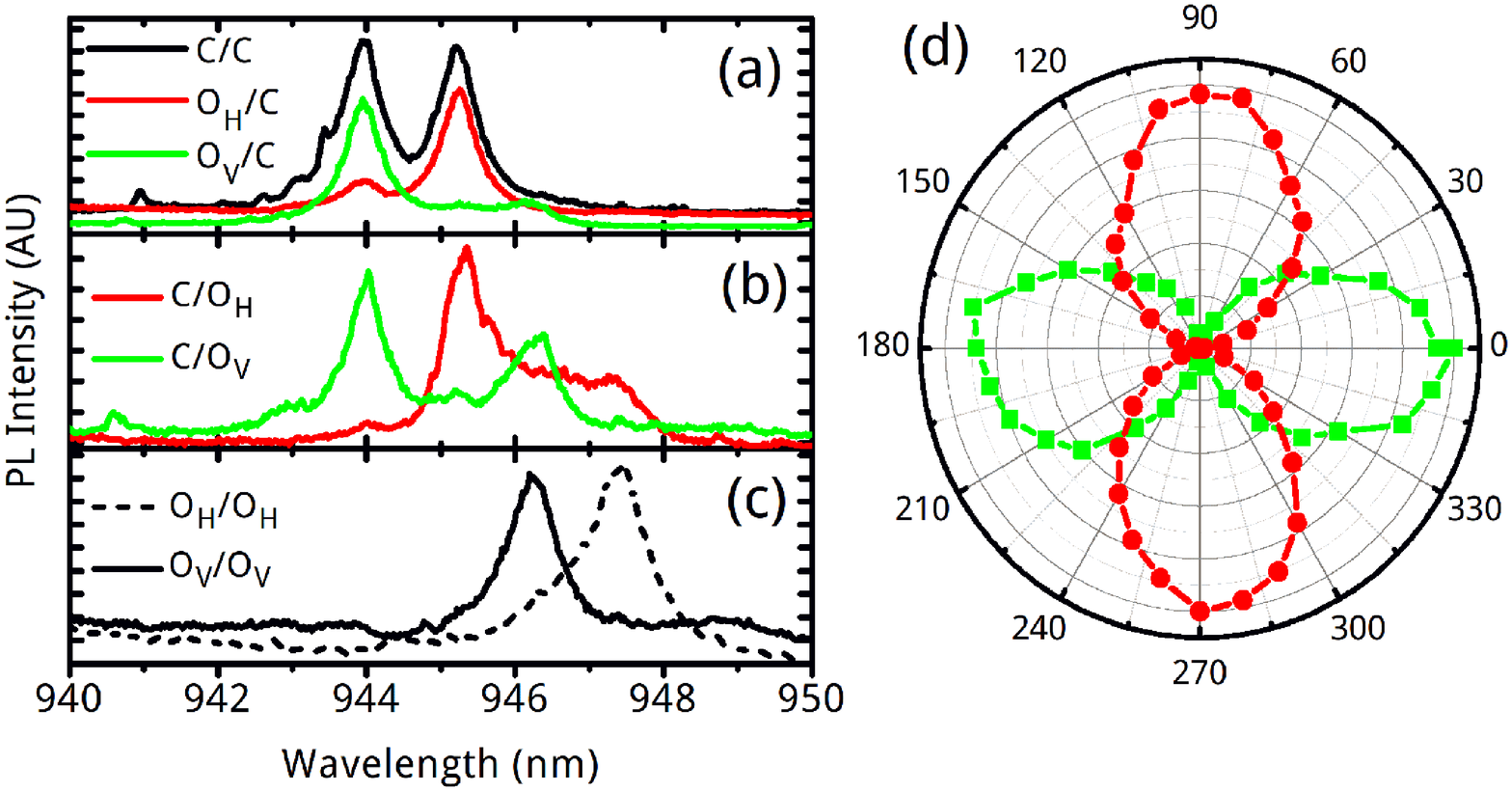}
\caption{ PL spectra obtained when (a) collecting from the cavity (b) exciting the outcouplers and collecting from the cavity (c) exciting and collecting from the vertical (solid black) and horizontal (dashed black) outcouplers. (d) polarization dependence of the two cavity modes when collecting from the cavity. The green curve (square markers) corresponds to the peak centerd at 943.9nm and red (circular markers) to the peak at 945.2nm.}
\label{nondegen_spectra}
\end{figure}

The quality factor of the cavity was first assessed by measuring the cavity modes using the C/C configuration. A typical coupled device spectrum is shown as the black line in Fig. \ref{nondegen_spectra}(a). Two cavity modes are clearly visible at 943.9nm and 945.2nm, with Q-factors of 1500 \& 1600 respectively. polarization-sensitive measurements of the cavity, shown in Fig. \ref{nondegen_spectra}(d), reveal that the modes are orthogonal: the peak at 943.9nm is horizontally (x) polarized and 945.2nm is vertically (y) polarized. To verify the selectivity of the cavity mode coupling to the waveguides, the C/O\textsubscript{H} \& C/O\textsubscript{V} configurations are used. As can be seen from Fig. \ref{nondegen_spectra}(b), the peak at 943.9nm is principally observed from O\textsubscript{V} and the 945.2nm peak from O\textsubscript{H}, confirming the behavior predicted by simulation.

The additional peaks observed at longer wavelengths (947.4nm for C/O\textsubscript{H} and 946.2nm for C/O\textsubscript{V}) are Fabry-Perot modes in the waveguides which are excited due to overlap with the cavity modes \cite{Sato2011}. This was determined by measuring both O\textsubscript{H}/O\textsubscript{H} and O\textsubscript{V}/O\textsubscript{V}, shown in Fig. \ref{nondegen_spectra}(c). In this geometry the Fabry-Perot modes of the waveguides are directly excited by QD PL and appear significantly brighter than the cavity modes. These Fabry-Perot resonances spectrally coincide with the features to longer wavelengths of the cavity modes as observed from the same waveguide in Fig. \ref{nondegen_spectra}(b).

Further proof of the selective coupling is observed in the O\textsubscript{H}/C and O\textsubscript{V}/C configurations, shown by the red and green curves in Fig.\ref{nondegen_spectra}(a). The selective coupling behavior is maintained and the cavity modes are excited with comparable intensity to direct excitation of the cavity (black curve). Due to the spectral filtering by the cavity, the Fabry-Perot modes in the waveguides are heavily suppressed in this geometry.

Measurements of uncoupled cavities yield typical Q-factors $\sim~$2400. These are much lower than the simulated Q-factors, implying significant disorder losses in the photonic crystal, leading to large scattering losses. Equation \ref{coup_eff_eq} implies comparable coupling efficiencies of 36\% \& 37\% for the X and Y-dipole modes respectively. These are much lower than the expected value of 89\% due to the large scattering loss rate of the cavities dominating over in-plane coupling rates in determining the total Q factor: this results in a small change in the Q-factor when the waveguides are introduced. Further improvements to fabrication are expected to improve the device efficiency by reducing these scattering losses, increasing the Q-factor of the uncoupled cavity.

The lifted degeneracy and low coupling efficiency of this cavity precludes its use as a spin-photon interface. However, since the coupling selection mechanisms rely upon spatial and not spectral discrimination, the selective coupling behavior of the cavity is maintained. Although some cross talk is observed, the background-subtracted contrast ratio ($\sim5$ for O\textsubscript{V} and $\sim10$ for O\textsubscript{H}) is sufficiently high to distinguish between the two. We believe the overlap of the FP mode in the Y-waveguide is responsible for the reduced contrast ratio, since the overlap region encompasses the X-dipole mode: when subtracted from the spectrum, this value increases to $\sim7$. Whilst unsuitable as a spin-photon interface, the non-degenerate cavity may be suitable for electro-optic switching applications. With application of a time-varying electric field \cite{Quilter2013,MichaelisdeVasconcellos2010,Faraon2010} to a single QD at the cavity center, the emission wavelength can be tuned from resonance with one mode to another via the quantum confined Stark effect. The emission rate into one mode is enhanced whilst the orthogonal component is suppressed \cite{Thijssen2012a} and the QD emission should be switched between the two waveguides on the timescale of the electric field modulation period.

\subsection{Near-Degenerate Cavity}

\begin{figure}[h]
\centering
\includegraphics[width=11cm]{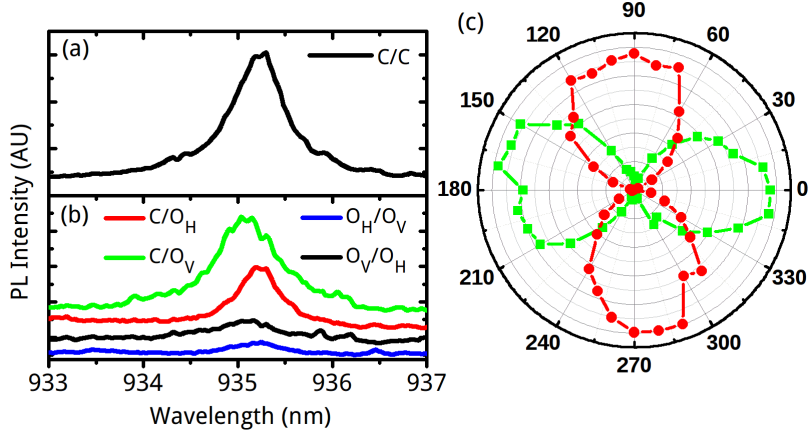}
\caption{PL spectra obtained when (a) exciting and collecting from the cavity (b) collecting from the outcouplers. In (b) the curves are offset for clarity. (c) polarization dependence of the two cavity modes when collecting from the cavity.}
\label{degen_spectra}
\end{figure}

Amongst the range of cavity mode splittings produced during fabrication, there are cavities which exhibit a mode splitting that is on the order of, or less than, the cavity mode linewidth. PL measurements on such a cavity are shown in Fig. \ref{degen_spectra}(a). The C/C configuration shows a single spectral feature centerd at 935.2nm. polarization measurements show this is comprised of two orthogonal modes as shown in Fig. \ref{degen_spectra}(c), with a splitting of 0.18nm. The Q-factors of the X(Y)-dipole modes are 1600(1200), with corresponding coupling efficiencies of 37(48)\%. This small splitting is reflected in the spectral measurements in Fig. \ref{degen_spectra}(b) for the C/O\textsubscript{H} and C/O\textsubscript{V} configuration. To assess the cross-talk, one outcoupler was excited whilst collecting from the other ((O\textsubscript{H}/O\textsubscript{V} and O\textsubscript{V}/O\textsubscript{H}), shown  in Fig. \ref{degen_spectra}(b): as can be seen from the data, the cross-talk is comparable to the background signal at $\sim$20\% of the peak intensity.

From these measurements, we conclude that this near-degenerate cavity also exhibits selective coupling of the orthogonal dipole modes to two separate waveguides, since the observed cross-talk is very low. Although this device also suffers from relatively low coupling efficiency, the Q factors remain sufficiently high to provide a high degree of spontaneous emission enhancement \cite{Laurent2005}. 

The out-of-plane emission of a single QD positioned at the cavity center which is spectrally resonant with the cavity modes would be unpolarized if orientated with the cavity modes, since it would couple to both dipole modes. We have demonstrated that these cavity modes couple to separate waveguides and hence only the corresponding orthogonally polarized components of the QD emission should be present in the waveguides; likewise, excitation via the waveguides will only couple the principle polarization components to the QD. This is similar to the result in \cite{Luxmoore2013}, with the possibility of spontaneous emission enhancement of the QD to produce indistinguishable single photons.

\section{Conclusion}
We have presented designs of an unpolarized photonic crystal cavity which exhibits selective coupling of the two orthogonally polarized dipole modes of an H1 cavity to two separate waveguides. Using FDTD simulations, the cavity-waveguide separation was optimized for equal coupling efficiencies of 89\%, coupled Q-factors exceeding 2000 and the waveguide dispersion adjusted to reduce propagation losses of the coupled cavity emission.

The selective coupling of the orthogonal dipole modes was experimentally demonstrated for a device with non-zero splitting of the cavity modes and for a device with a small splitting to linewidth ratio. The former may offer functionality as an electro-optic switch; the latter is expected to act as a spin-photon interface for a resonant QD positioned at the cavity center. This provides a one-to-one correspondence between the Bloch sphere of the excitonic spin states of the QD and the Poincar\'{e}-like sphere of the cavity modes, encoding this information in a which-path regime. The device maintains selectivity when excited via either the cavity or the waveguides, such that several devices coupled together may realize a scalable quantum spin network, although single QD measurements remain to fully confirm these predictions.

The authors thank D.M. Whittaker, A.J. Ramsay and N.A. Wasley for helpful discussions. This work was funded by the EPSRC research grant number: EP/J007544/1.

\end{document}